\documentclass[aps,showpacs,twocolumn,pra,floatfix]{revtex4}
\usepackage{amssymb}
\usepackage{bm}
\usepackage{graphicx}
\usepackage[latin1]{inputenc}
\begin{document}
\title{Families of bipartite states classifiable by the positive partial transposition criterion}
\author{F. E. S. Steinhoff}
\email{steinhof@ifi.unicamp.br}
\author{M. C. de Oliveira}
\email{marcos@ifi.unicamp.br}
\affiliation{ Instituto de F\'\i sica ``Gleb Wataghin'', Universidade Estadual de Campinas -UNICAMP,\\
13083-970, Campinas, SP, Brazil}
\begin{abstract}
We construct a family of bipartite states of arbitrary dimension whose eigenvalues of the partially transposed matrix can be inferred directly from the block structure of the global density matrix. We identify from this several subfamilies in which the PPT criterion is both necessary and sufficient. A sufficient criterion of separability is obtained, which is fundamental for the discussion. We show how several examples of states known to be classifiable by the PPT criterion indeed belong to this general set. Possible uses of these states in numerical analysis of entanglement and in the search of PPT bound entangled states are briefly discussed.
\end{abstract}

\maketitle
\section{Introduction}

Quantum entanglement has a major role in nowadays discussions about quantum information processing due to its potential application in protocols \cite{nielsen}. Despite its importance, entanglement characterization has been recognized as a difficult task. %Even to say if a state is entangled or not is a very complex problem \cite{gurvits}. 
Such difficulties motivates the search for alternative ways of detecting the separability of a given state. Operational separability criteria based on positive, but not completely positive, maps appeared \cite{peres,horodecki1,horodecki2,rudolph}, with relative success. 
In the bipartite case, the most important of these criteria is the Positivity under Partial Transposition (PPT) criterion, due to Peres \cite{peres}. It asserts that if a state is separable, then its partial transposition will be a positive semidefinite operator. By partial transposition it is understood the operation of transposing the matrix elements of only one of the subsystems. It is sometimes referred as partial specular reflection operation or local time reversal operation \cite{sanpera}.
A very illustrative way of seeing partial transposition operation is by considering the density matrix of the state in the basis $\{|0,0\rangle,|0,1\rangle,\ldots, |0,d_B-1\rangle, |1,0\rangle, |1,1\rangle, \ldots, |d_A-1,d_B-1\rangle\}$ - called here Standard Computational Basis (SCB) - where $d_A$ and $d_B$ are the dimensions of Alice and Bob´s subsystems, respectively,
\begin{eqnarray}
\rho = \left(\begin{array}{c c c}{A_{00}}&{\ldots}&{A_{0,d_A-1}}\\{\vdots}&
{\ddots}&{\vdots}\\{A_{0,d_A-1}^{\dagger}}&{\ldots}&{A_{d_A-1,d_A-1}}\end{array}\right),
\label{state1}\end{eqnarray}
with $A_{ij}$ being $d_B\times d_B$ submatrices. The partial transposition of the state (\ref{state1}) is simply
\begin{eqnarray}
\rho^{\Gamma} = \left(\begin{array}{c c c}{(A_{00})^T}&{\ldots}&{(A_{0,d_A-1})^T}\\{\vdots}&
{\ddots}&{\vdots}\\{(A_{0,d_A-1}^{\dagger})^T}&{\ldots}&{(A_{d_A-1,d_A-1})^T}\end{array}\right),\label{nice}
\end{eqnarray}
where we remark the importance of the ordering of the basis. In a different ordering the partially transposed matrix would assume a different aspect. If one has a separable state $\rho$, then PPT criterion assures that $\rho^{\Gamma}$ will be a positive semidefinite operator. But the converse is not generally true, making PPT criterion only a necessary one. In fact, PPT criterion has been shown to be necessary and sufficient for some special classes of states: two-qubit and qubit-qutrit states \cite{horodecki1}, Werner \cite{werner} and isotropic states \cite{horodecki2}, low-rank states \cite{horodecki3} and, in a different way, Gaussian states in continuous-variable context \cite{simon}. Other states which are positive under partial transposition, but are known, by other means, to be entangled belong to the class of bound entangled states \cite{horodecki2}. Those states have no distillable entanglement, i. e., no pure entangled states can be extracted  through local operations and classical communication.

Among the open problems in quantum information theory, one which is extremely important is to encounter general positive but not completely positive maps that would detect PPT entangled states for systems with arbitrary Hilbert space dimension. Another important direction to follow is to search for classes of states for which the PPT criterion is a sufficient one for $d_A.d_B > 6$. Recent progress was given by \cite{kossakowski} and references therein, where classes of PPT states were obtained. %Also, an important question that can be posed is whether there are states, which are not positive under partial transposition and also bound entangled. 

The aim of this paper is to present novel families of bipartite states of arbitrary dimension whose eigenvalues of $\rho^{\Gamma}$ are easily inferred from the block structure of the state, allowing one to tell if the state is PPT and to compute its negativity straightforwardly. For several subfamilies we will see that PPT criterion is both necessary and sufficient. These subfamilies include several examples cited above, and known to follow the PPT criterion. With this extension of the set of states classifiable through the PPT criterion we expect some advantages for numerical analysis of entanglement as well as for discussions about bound entanglement. The paper is divided as follows: in Sec. II we present an illustrative example, in order to motivate the discussion. In Sec. III we present the family of states for arbitrary $d_A$ and $d_B$. In Sec. IV we prove a simple sufficient separability condition, identifying with this some important subfamilies in which positivity under partial transposition is equivalent to separability; in further subsections, we obtain nontrivial decompositions of the state space into direct-sums, which raise interesting questions and we also obtain yet another PPT-classifiable set of states. Finally, in Sec. V we present our conclusions, enclosing the paper. 

\section{First example: $d_A=d_B=4$}
We start our discussion by presenting an illustrative example, having all the characteristics we want to generalize.
We analyze first a situation in which Alice and Bob´s subsystems are four dimensional.  Consider then the following matrix in the SCB,
\begin{widetext}{\begin{eqnarray}
\rho = \left(\begin{array}{c c c c | c c c c | c c c c | c c c c}
          {x_{00}}&{0}&{0}&{0}&{0}&{0}&{0}&{0}&{0}&{0}&{0}&{0}&{0}&{0}&{0}&{0} 
\\        {0}&{a_1}&{a_4}&{a_6}&{x_{01}}&{0}&{0}&{0}&{0}&{0}&{0}&{0}&{0}&{0}&{0}&{0} 
\\        {0}&{a_4^*}&{a_2}&{a_5}&{0}&{0}&{0}&{0}&{x_{02}}&{0}&{0}&{0}&{0}&{0}&{0}&{0}
\\        {0}&{a_6^*}&{a_5^*}&{a_3}&{0}&{0}&{0}&{0}&{0}&{0}&{0}&{0}&{x_{03}}&{0}&{0}&{0}
\\ \hline {0}&{x_{01}^*}&{0}&{0}&{b_1}&{0}&{0}&{0}&{0}&{0}&{0}&{0}&{0}&{0}&{0}&{0}
\\        {0}&{0}&{0}&{0}&{0}&{x_{11}}&{0}&{0}&{0}&{0}&{0}&{0}&{0}&{0}&{0}&{0}
\\        {0}&{0}&{0}&{0}&{0}&{0}&{b_2}&{b_4}&{0}&{x_{12}}&{0}&{0}&{0}&{0}&{0}&{0}
\\        {0}&{0}&{0}&{0}&{0}&{0}&{b_4^*}&{b_3}&{0}&{0}&{0}&{0}&{0}&{x_{13}}&{0}&{0}
\\ \hline {0}&{0}&{x_{02}^*}&{0}&{0}&{0}&{0}&{0}&{c_1}&{c_4}&{0}&{0}&{0}&{0}&{0}&{0}
\\        {0}&{0}&{0}&{0}&{0}&{0}&{x_{12}^*}&{0}&{c_4^*}&{c_2}&{0}&{0}&{0}&{0}&{0}&{0}
\\        {0}&{0}&{0}&{0}&{0}&{0}&{0}&{0}&{0}&{0}&{x_{22}}&{0}&{0}&{0}&{0}&{0}
\\        {0}&{0}&{0}&{0}&{0}&{0}&{0}&{0}&{0}&{0}&{0}&{c_3}&{0}&{0}&{x_{23}}&{0}
\\ \hline {0}&{0}&{0}&{x_{03}^*}&{0}&{0}&{0}&{0}&{0}&{0}&{0}&{0}&{d_1}&{d_4}&{d_6}&{0}
\\        {0}&{0}&{0}&{0}&{0}&{0}&{0}&{x_{13}^*}&{0}&{0}&{0}&{0}&{d_4^*}&{d_2}&{d_5}&{0}
\\        {0}&{0}&{0}&{0}&{0}&{0}&{0}&{0}&{0}&{0}&{0}&{x_{23}^*}&{d_6^*}&{d_5^*}&{d_3}&{0}
\\        {0}&{0}&{0}&{0}&{0}&{0}&{0}&{0}&{0}&{0}&{0}&{0}&{0}&{0}&{0}&{x_{33}}
\end{array}\right),\label{m1}
\end{eqnarray}}
%\end{widetext}
which under partial transposition on Bob's subsystem writes as
%\begin{widetext}
\begin{eqnarray}
\rho^{\Gamma} = \left(\begin{array}{c c c c | c c c c | c c c c | c c c c}
          {x_{00}}&{0}&{0}&{0}&{0}&{x_{01}}&{0}&{0}&{0}&{0}&{x_{02}}&{0}&{0}&{0}&{0}&{x_{03}} 
\\        {0}&{a_1}&{a_4^*}&{a_6^*}&{0}&{0}&{0}&{0}&{0}&{0}&{0}&{0}&{0}&{0}&{0}&{0} 
\\        {0}&{a_4}&{a_2}&{a_5^*}&{0}&{0}&{0}&{0}&{0}&{0}&{0}&{0}&{0}&{0}&{0}&{0}
\\        {0}&{a_6}&{a_5}&{a_3}&{0}&{0}&{0}&{0}&{0}&{0}&{0}&{0}&{0}&{0}&{0}&{0}
\\ \hline {0}&{0}&{0}&{0}&{b_1}&{0}&{0}&{0}&{0}&{0}&{0}&{0}&{0}&{0}&{0}&{0}
\\        {x_{01}^*}&{0}&{0}&{0}&{0}&{x_{11}}&{0}&{0}&{0}&{0}&{x_{12}}&{0}&{0}&{0}&{0}&{x_{13}}
\\        {0}&{0}&{0}&{0}&{0}&{0}&{b_2}&{b_4^*}&{0}&{0}&{0}&{0}&{0}&{0}&{0}&{0}
\\        {0}&{0}&{0}&{0}&{0}&{0}&{b_4}&{b_3}&{0}&{0}&{0}&{0}&{0}&{0}&{0}&{0}
\\ \hline {0}&{0}&{0}&{0}&{0}&{0}&{0}&{0}&{c_1}&{c_4^*}&{0}&{0}&{0}&{0}&{0}&{0}
\\        {0}&{0}&{0}&{0}&{0}&{0}&{0}&{0}&{c_4}&{c_2}&{0}&{0}&{0}&{0}&{0}&{0}
\\        {x_{02}^*}&{0}&{0}&{0}&{0}&{x_{12}^*}&{0}&{0}&{0}&{0}&{x_{22}}&{0}&{0}&{0}&{0}&{x_{23}}
\\        {0}&{0}&{0}&{0}&{0}&{0}&{0}&{0}&{0}&{0}&{0}&{c_3}&{0}&{0}&{0}&{0}
\\ \hline {0}&{0}&{0}&{0}&{0}&{0}&{0}&{0}&{0}&{0}&{0}&{0}&{d_1}&{d_4^*}&{d_6^*}&{0}
\\        {0}&{0}&{0}&{0}&{0}&{0}&{0}&{0}&{0}&{0}&{0}&{0}&{d_4}&{d_2}&{d_5^*}&{0}
\\        {0}&{0}&{0}&{0}&{0}&{0}&{0}&{0}&{0}&{0}&{0}&{0}&{d_6}&{d_5}&{d_3}&{0}
\\        {x_{03}^*}&{0}&{0}&{0}&{0}&{x_{13}^*}&{0}&{0}&{0}&{0}&{x_{23}^*}&{0}&{0}&{0}&{0}&{x_{33}}
\end{array}\right). \label{m2}
\end{eqnarray}
\end{widetext}
It is evident that for the matrix (\ref{m1}) to represent a valid state, it must be normalised with $x_{00}+\sum_{i=1}^3\left(x_{ii}+a_i+b_i+c_i+d_i\right)=1$ and $\rho$ must be positive semidefinite. It is important here to note that this latter condition implies that the submatrices 
\begin{eqnarray}
A = \left(\begin{array} {c c c} {a_1}&{a_4}&{a_6} \\ {a_4^*}&{a_{2}}&{a_5} \\ {a_6^*}&{a_5^*}&{a_3} \end{array}\right); \ \ B = \left(\begin{array} {c c} {b_2}&{b_4} \\ {b_4^*}&{b_3} \end{array}\right);\nonumber\\  C = \left(\begin{array} {c c} {c_1}&{c_4} \\ {c_4^*}&{c_2} \end{array}\right); \ \ D = \left(\begin{array} {c c c} {d_1}&{d_4}&{d_6} \\ {d_4^*}&{d_{2}}&{d_5} \\ {d_6^*}&{d_5^*}&{d_3} \end{array}\right), \label{set1}
\end{eqnarray}
are also positive semidefinite, since they are principal submatrices of $\rho$ \cite{horn}. The alternating sizes of these blocks are also a fundamental property for the extension for arbitrary dimension. 

We claim that the operator defined by (\ref{m2}) has a direct-sum structure
\begin{eqnarray}
\rho^{\Gamma}=X\oplus A^T\oplus b_1\oplus B^T\oplus C^T\oplus c_3\oplus D^T, \label{decomposition}
\end{eqnarray}
where 
\begin{eqnarray}
X=\left(\begin{array}{c c c c}{x_{00}}&{x_{01}}&{x_{02}}&{x_{03}}\\{x_{01}^*}&{x_{11}}&{x_{12}}&{x_{13}}\\{x_{02}^*}&{x_{12}^*}&{x_{22}}&{x_{23}}\\{x_{03}^*}&{x_{13}^*}&{x_{23}^*}&{x_{33}}\end{array}\right).\label{set2}
\end{eqnarray}
To see this, note that we can decompose the total Hilbert space $\mathcal{H}$ as a direct sum:
\begin{eqnarray}
\mathcal{H}=\mathcal{H}_X\oplus\mathcal{H}_A\oplus\mathcal{H}_{b_1}\oplus\mathcal{H}_B\oplus\mathcal{H}_C\oplus\mathcal{H}_{c_3}\oplus\mathcal{H}_D
\end{eqnarray}
with support in the orthogonal subspaces
\begin{eqnarray}
\mathcal{H}_X &=& span\{|00\rangle,|11\rangle,|22\rangle,|33\rangle\}; \label{sb1} \\ \mathcal{H}_A &=& span\{|01\rangle,|02\rangle,|03\rangle\}; \\ \mathcal{H}_{b_1} &=& span\{|10\rangle\}; \\ \mathcal{H}_B &=& span\{|12\rangle,|13\rangle\}; \\ \mathcal{H}_C &=& span\{|20\rangle, |21\rangle\}; \\ \mathcal{H}_{c_3} &=& span\{|23\rangle\}; \\ \mathcal{H}_D &=& span\{|30\rangle,|31\rangle,|32\rangle\}. \label{subspaces}
\end{eqnarray}
Since the matrices (\ref{set1}) and (\ref{set2}) have support only in the subspaces with respective indexes in (\ref{sb1})-(\ref{subspaces}), then we have the decomposition (\ref{decomposition}) for partial transposition. In the Appendix A we consider this in detail working out an example. We note that similar decompositions are given in \cite{kossakowski}, where the authors construct several classes of PPT states decomposing the total Hilbert space $\mathcal{H}$ into well-suited direct sums, which can be changed through circular property of the supports. 

The eigenvalues of $\rho^{\Gamma}$ are thus $b_1$, $c_3$ and the eigenvalues of the matrices $X$, $A^T$, $B^T$, $C^T$ and $D^T$. Since $A$, $B$, $C$ and $D$ are positive semidefinite - this is implied by the positive semidefiniteness of $\rho$ - so are $A^T$, $B^T$, $C^T$ and $D^T$. Of course $b_1,c_3\geq 0$ and then we conclude that the only negative eigenvalues of $\rho^{\Gamma}$, if any, are the negative eigenvalues of the matrix $X$. So, if the $X$ matrix has any negative eigenvalue, we can say that the state is entangled, by PPT criterion. Also, its Negativity \cite{vidal} will be simply the sum of these negative eigenvalues. However, if $\rho^{\Gamma}$ is positive semidefinite, in general we cannot say if $\rho$ in Eq. (\ref{m1}) is separable or not. But by adding some constraints we restrict $\rho$ to a separable state. In the next section, devoted to the general case, we see how it works.

\section{Generalization to arbitrary $d_A$, $d_B$}

The generalization of the previous example to arbitrary dimensions is not a difficult task. We are assuming here that $d_A\leq d_B$. The expression of an arbitrary member of the family expressed in terms of the SCB $\{|0,0\rangle,|0,1\rangle,\ldots, |0,d_B-1\rangle, |1,0\rangle, |1,1\rangle, \ldots, |d_A-1,d_B-1\rangle\}$ is given by
\begin{eqnarray}
\rho &=& \sum_{m,n=0}^{d_A-1}x_{mn}|mn\rangle\langle nm|+\sum_{k=0}^{d_A-1}|k\rangle\langle k|\label{final}\\
&&\otimes\left(\sum_{i,j=0}^{k-1}(M_{k})_{ij}|i\rangle\langle j|
+\sum_{i',j'=k+1}^{d_B-1}(N_{k})_{i'j'}|i'\rangle\langle j'|\right)\nonumber. 
\end{eqnarray}
We impose on this operator the following conditions:
\begin{enumerate}
	\item $(M_{k})_{ji}=(M_{k})_{ij}^*$, $(N_{k})_{ji}=(N_{k})_{ij}^*$, $x_{nm}=x_{mn}^*$;
	\item $\sum_{k=0}^{d_A-1}\left(\sum_{i=0}^{k-1}(M_{k})_{ii}+\sum_{i=k+1}^{d_B-1}(M_{k})_{ii}\right)\\+\sum_{m=0}^{d_A-1} x_{mm}=1$;
	\item $\rho$ is positive semidefinite.
\end{enumerate}
These are the usual conditions to be fulfilled for $\rho$ to represent a valid state, that is, hermiticity, unit trace and positive semidefiniteness, respectively. This is easier to see if one consider the block structure of $\rho$ in the SCB. Any operator can be expressed in this basis as
\begin{eqnarray}
\rho = \left(\begin{array}{c c c}{A_{00}}&{\ldots}&{A_{0,d_A-1}}\\{\vdots}&
{\ddots}&{\vdots}\\{A_{0,d_A-1}^{\dagger}}&{\ldots}&{A_{d_A-1,d_A-1}}\end{array}\right),
\end{eqnarray}
where $A_{ij}$ are $d_B\times d_B$ submatrices. The states just constructed have diagonal submatrices given by
\begin{eqnarray}
A_{kk} = \left(\begin{array} {c | c | c} {(M_k)_{k\times k}}&{}&{} \\ \hline {}&{x_{kk}}&{} \\ \hline {}&{}&{(N_k)_{(d_B-1-k)\times (d_B-1-k)}}\end{array}\right),
\end{eqnarray}
with $k=0,1,\ldots,d_A-1$, $M_k$ and $N_k$ being diagonal blocks of dimensions given by the respective subindexes and $x_{kk}$ is an arbitrary real number. The off-diagonal submatrices are simply $A_{ij}=x_{ij}|j\rangle\langle i|$, with $i\neq j$. The conditions on the elements in this fashion are simply that $M_k$, $N_k$ and the matrix with elements $x_{ij}$ are Hermitian, that the sum of their traces sum up to unit and that the global matrix is positive semidefinite. 

Let $\rho^{\Gamma}$ be the operator obtained from $\rho$ through the partial transposition of Bob´s subsystem:
\begin{eqnarray}
\rho^{\Gamma} &=&  \sum_{m,n=0}^{d_A-1}x_{mn}|mm\rangle\langle nn|+\sum_{k=0}^{d_A-1}|k\rangle\langle k|\\
&&\otimes\left(\sum_{i,j=0}^{k-1}(M_{k})_{ij}|j\rangle\langle i|
%\right.\\&&+\left.
+\sum_{i',j'=k+1}^{d_B-1}(N_{k})_{i'j'}|j'\rangle\langle i'|\right).\nonumber
\end{eqnarray} 
The first term corresponds to a $d_B\times d_B$ matrix - called $X$ here - acting in the subspace spanned by $\{|00\rangle, |11\rangle, \ldots, |d_A-1, d_A-1\}$ only. The matrices $M_k^T$ act in the subspace spanned by $\{|k,0\rangle,|k,1\rangle,\ldots,|k,k-1\rangle\}$ only, while the matrices $N_k^T$ act in the subspace spanned by $\{|k,k+1\rangle,|k,k+2\rangle.\ldots,|k,d_B-1\rangle\}$ only. As the intersection between every two of these subspaces is the null vector, the total Hilbert space $\mathcal{H}$ can be decomposed as a direct sum of them and by the reasoning given, the operator $\rho^{\Gamma}$ has a direct sum structure, which can be compactly stated as
\begin{eqnarray} 
\rho^{\Gamma} = X\bigoplus_{k=0}^{d_A-1}\left(M_k^T\oplus N_k^T\right). \label{magic}
\end{eqnarray}
The eigenvalues of the transposed matrix are thus the eigenvalues of the various matrices $M_k$, $N_k$ and $X$. However, a negative eigenvalue of $\rho^{\Gamma}$, if any, will be due only to a negative eigenvalue of the $X$ matrix. The reason is that $M_k$ and $N_k$ are principal submatrices of the original density matrix $\rho$. By elementary linear algebra \cite{horn}, these matrices are already positive semidefinite, since $\rho$ is positive semidefinite. So, the only negative eigenvalues of $\rho^{\Gamma}$ will be the negative eigenvalues of the $X$ matrix.

The $X$ matrix has thus a large part of the information about the entanglement of the state. If the states just constructed have any experimental usage, to detect and to quantify the entanglement of $\rho$ will be simpler than full state reconstruction. The $X$ matrix is a $d_A\times d_A$ matrix and even if one has to reconstruct this matrix  \footnote{It would be highly desirable that the eigenvalues of $X$ could be inferred from simple experimental schemes.} this task will be much simpler than to reconstruct the $(d_A.d_B)\times(d_A.d_B)$ global matrix representing $\rho$. In practice, one should know that the state is of this form, by the preparation procedure or by some characteristic - and yet not discovered - test. Also, from another point of view, we believe that these states would bring advantage in numerical research of entanglement, since one has to deal only with one matrix. It is easy to construct entangled states with the form (\ref{final}) and, as will be shown in the next section, it is also easy to construct families of PPT-classifiable states.  

However, if the measured eigenvalues of $X$ are all positive, then the PPT criterion alone will not be sufficient to detect entanglement, once in the general bipartite case this criterion is only necessary for separability. As the state is PPT, if it is shown to be entangled by other means, it will exhibit bound entanglement \cite{horodecki2}. 

\section{Subfamilies classifiable through PPT criterion}

There are subfamilies inside the broad family of bipartite states presented in the last section in which PPT criterion is necessary and sufficient. We do not intend to present here all the situations, but instead to discuss how it includes some very important examples. For that, we would like first to prove a simple and relevant result:

\newtheorem{prop}{Proposition}
\begin{prop}
If a state $\rho$ expressed in the standard computational basis has the block diagonal form
\begin{eqnarray}
\rho_{ss} = \left(\begin{array}{c | c | c | c}
{A_{0}}&{}&{}&{} \\ \hline {}&{A_{1}}&{}&{} \\ \hline {}&{}&{\ddots}&{} \\ \hline{}&{}&{}&{A_{d-1}}\end{array}\right)\label{ss},
\end{eqnarray}
where each $A_i$ is a $d_B\times d_B$ matrix, then the state is separable.
\end{prop}
\textit{Proof:} We have to prove that a matrix in the form $\rho_{ss}$ above has a decomposition
\begin{eqnarray}
\rho_s = \sum_i p_i \rho_A^i\otimes\rho_B^i,
\end{eqnarray}
with $\sum_i p_i =1$, $p_i\geq 0$ and $\rho_A^i$, $\rho_B^i$ being states in Alice and Bob´s subsystems, respectively. Indeed, we have
\begin{eqnarray}
\rho_{ss} &=& |0\rangle\langle 0|\otimes A_0 + |1\rangle\langle 1|\otimes A_1 + \ldots \nonumber\\
&&+ |d-1\rangle\langle d-1|\otimes A_{d-1} = \sum_{i=0}^{d-1} |i\rangle\langle i|\otimes A_i,\end{eqnarray} which thus can be written as \begin{eqnarray}\rho_{ss} = \sum_{i=0}^{d-1} \underbrace{(trA_i)}_{p_i}\underbrace{|i\rangle\langle i|}_{\rho_A^i}\otimes\underbrace{\frac{A_i}{trA_i}}_{\rho_B^i} = \sum_i p_i \rho_A^i\otimes\rho_B^i,
\end{eqnarray}
and since $\sum_i p_i = \sum_i trA_i = 1$ and $p_i\geq 0$, a state in the form $\rho_{ss}$ is separable. \hfill\rule{3mm}{3mm} \\

We will use this result as a probe to construct subfamilies of states classifiable through PPT criterion. We will call states that can be written in the block diagonal form (\ref{ss}) as \textit{simply separable states}. It is straightforward then that every state written in the SCB can be decomposed as $\rho=\rho_{ss}+M$, i.e., a simply separable state $\rho_{ss}$ plus a matrix $M$ that do not represent a state and that contains correlations associated with entanglement. We show that important examples, such as the Werner states are included in this subset.

\subsection{Restrictions on the $X$ matrix}

We shall construct examples of states in which PPT means separability restricting the form of the $X$ matrix appearing in the direct-sum decomposition of $\rho^{\Gamma}$. If, for example, the eigenvalues of the $X$ matrix are all of the form $\{-x_i\}_{i=0}^{d_A-1}$, for arbitrary real values $x_i$, then the state will have a positive partial transposition only in case all the $x_i$ are zero. But in this case the $X$ matrix can only be the null matrix, which implies that the matrix $\rho$ is block diagonal with $d_B\times d_B$ blocks, that is, the state is simply separable (Proposition 1). In this special case, positivity under partial transposition implies separability, i. e., we obtain a subfamily in which states are separable if and only if they are PPT \footnote{That separability implies positivity under partial transposition is already true, by the PPT criterion.}. Another example can be given for a $X$ matrix of the form ($d_A$ even)
\begin{eqnarray}
X=\bigoplus_{i=0}^{d_A/2-1}\left(\begin{array}{c c}{0}&{x_i}\\{x_i^*}&{0}\end{array}\right).
\end{eqnarray}
The eigenvalues of this matrix are obviously $\{\pm |x_i|\}_{i=0}^{d_A/2-1}$. The matrix $\rho^{\Gamma}$ will be positive semidefinite if and only if all the $x_i$ are zero. This implies that $\rho$ is simply separable, i. e., we have again equivalence between separable and PPT states in this case. Indeed, whenever positivity under partial transposition implies that the $X$ matrix is the null matrix we will have this equivalence.

We obtain another such  a subfamily whenever the $X$ matrix is itself diagonal. In this case it is clear that $\rho$ will be simply separable. Combining this with the previous reasoning, whenever positivity under partial transposition implies that the $X$ matrix is diagonal, we will have equivalence between separable and PPT states. In fact, the previous case can be trivially seen as a special case of this one. Consider, for example, a $X$ matrix of the form
\begin{eqnarray}
X=\bigoplus_{i=0}^{d_A/2-1}\left(\begin{array}{c c}{0}&{x_i}\\{x_i^*}&{y_i}\end{array}\right).
\end{eqnarray}
This matrix will be positive if and only if $x_i=0$; in this case, $X$ will be diagonal and the state will be simply separable. We showed thus several examples of subfamilies classifiable through PPT criterion. 

\subsection{Werner and isotropic states}

For a $d\otimes d$ system, consider the state 
\begin{eqnarray}
\rho_W = (1-\epsilon)\frac{I}{d^2} + \epsilon\frac{F}{d} \label{wstate},
\end{eqnarray}
where $I$ is the identity operator and $F$ is the usual flip operator, defined by $F|\phi\rangle\otimes |\varphi\rangle=|\varphi\rangle\otimes |\phi\rangle$. We call the family of states defined by (\ref{wstate}) as Werner states, \cite{werner}; the connection with Werner´s original notation is $\epsilon=-\frac{1-d\Phi}{d^2 -1}$, where $\Phi=\langle F\rangle$. If we impose that the $M_k$, $N_k$ matrices are of the form $\frac{(1-\epsilon)}{d^2}I$ and the $X$ matrix elements are $x_{kk}=(1-\epsilon(d+1))/d^2$ and $x_{jk}=\epsilon/d$, for $j\neq k$, then we see that Werner states are also a subfamily of the broader family. A Werner state is separable if and only if it is PPT, as is well known. 

The partial transposition establishes a nice connection between Werner and the so called isotropic states \cite{horodecki2} given by
\begin{eqnarray}
\rho_I = (1-\epsilon)\frac{I_A\otimes I_B}{d^2} +\epsilon P_+ \label{isotr}
\end{eqnarray}
where $P_+=|\phi_d^+\rangle\langle\phi_d^+|$ and $|\phi_d^+\rangle = \frac{1}{\sqrt{d}}\sum_{i=0}^{d-1}|ii\rangle$. Since the partial tranposition of the operator $P_+$ is simply $F/d$, we have that an isotropic state will be PPT only if its partial tranposition $\rho_I^{\Gamma}$ represents a Werner state and the same statement holds for a Werner state. However, the isotropic states do not belong to the family defined by (\ref{final}). But we can easily define a new family which contains isotropic states, constituted of matrices in the form (\ref{magic}) and restricting the submatrices $M_k$ and $N_k$ to be diagonal. Making $M_k$ and $N_k$ equal to $\frac{(1-\epsilon)}{d^2}I$ and $x_{kk}=(1-\epsilon(d+1))/d^2$, $x_{jk}=\epsilon/d$, for $j\neq k$, we see that isotropic states are contained in this new family. We get then an analogous connection between the two broader subfamilies through the partial transposition operation. 

\subsection{Qubit-qubit, qubit-qutrit and qubit-qudit cases}

The PPT criterion is necessary and sufficient for $d_A=d_B=2$ and $d_A=2, d_B=3$ \cite{horodecki1}. The density matrix of a two qubit state in the SCB of the family reads
\begin{eqnarray}
\rho = \left(\begin{array}{c c | c c} {x_{00}}&{0}&{0}&{0}\\{0}&{a}&{x_{01}}&{0}\\ \hline {0}&{x_{01}^*}&{b}&{0}\\{0}&{0}&{0}&{x_{11}}\end{array}\right),
\end{eqnarray}
and its partial transposition is
\begin{eqnarray}
\rho^{\Gamma} = \left(\begin{array}{c c | c c} {x_{00}}&{0}&{0}&{x_{01}}\\{0}&{a}&{0}&{0}\\ \hline {0}&{0}&{b}&{0}\\{x_{01}^*}&{0}&{0}&{x_{11}}\end{array}\right).
\end{eqnarray}
The eigenvalues of $\rho^{\Gamma}$ are $a$, $b$ and the eigenvalues of the matrix
\begin{eqnarray}
X= \left(\begin{array}{c c} {x_{00}}&{x_{01}} \\ {x_{01}^*}&{x_{11}} \end{array}\right),
\end{eqnarray}
which are simply $\frac{1}{2}\left(x_{00}+x_{11}\pm\sqrt{(x_{00}-x_{11})^2 + 4|x_{01}|^2}\right)$. The state will be PPT and hence separable for $x_{00}x_{11}\geq |x_{01}|^2$. Otherwise, the state will be entangled and its negativity will be given by $N(\rho)=\frac{1}{2} max\{0,\sqrt{(x_{00}-x_{11})^2 + 4|x_{01}|^2}-(x_{00}+x_{11})\}$.

In case we are dealing with a qubit and a qutrit, the density matrix of the family reads
\begin{eqnarray}
\rho = \left(\begin{array}{c c c | c c c} 
{x_{00}}&{0}&{0}&{0}&{0}&{0} \\
{0}&{a}&{c}&{x_{01}}&{0}&{0} \\
{0}&{c^*}&{b}&{0}&{0}&{0} \\ \hline
{0}&{x_{01}^*}&{0}&{d}&{0}&{0} \\
{0}&{0}&{0}&{0}&{x_{11}}&{0} \\
{0}&{0}&{0}&{0}&{0}&{e} \end{array}\right),
\end{eqnarray}
and is easy to see that the same results apply to this case. In fact, from a more general density matrix
\begin{eqnarray}
\rho = \left(\begin{array}{c c c | c c c} 
{x_{00}}&{0}&{0}&{0}&{0}&{0} \\
{0}&{a}&{c}&{0}&{0}&{0} \\
{0}&{c^*}&{b}&{x_{01}}&{0}&{0} \\ \hline
{0}&{0}&{x_{01}^*}&{d}&{f}&{0} \\
{0}&{0}&{0}&{f^*}&{e}&{0} \\
{0}&{0}&{0}&{0}&{0}&{x_{11}} \end{array}\right),
\end{eqnarray}
we see that the negative eigenvalues of $\rho^{\Gamma}$ are also equal to the cases analyzed. We are thus induced to propose another family of states for the qubit-qudit case given by
\begin{eqnarray}
\rho &=& x_{00}|00\rangle\langle 00| + x_{11}|d_B-1,d_B-1\rangle\langle d_B-1,d_B-1| \nonumber\\&&+ x_{01}|0,d_B-1\rangle\langle 10| + x_{01}^*|10\rangle\langle 0,d_B-1| \nonumber\\ 
&&+ \sum_{i,j=1}^{d_B-1}A_{ij}|0i\rangle\langle 0j|+ \sum_{i'j'=0}^{d_B-2}B_{i'j'}|1i'\rangle\langle 1j'|, \label{2xN}
\end{eqnarray}
and it is easy to see that the negative eigenvalues of the transposed matrix will be the same. However, in this case it is not assured that positivity under partial transposition implies separability. One can conjecture here that this is indeed true, given the resemblance to the qubit-qubit and qutrit-qutrit cases. We make here a brief digression about this subject and we hope it can be useful for the search of bound entangled states. The partial transposes of (\ref{final}) and (\ref{2xN}) are of the form 
\begin{eqnarray}
\rho^{\Gamma} = X\oplus\tilde{\rho}_{ss} \label{ptrans}
\end{eqnarray}
where $\tilde{\rho}_{ss}$ is an unnormalised simply separable density operator. As we are looking for bound entangled states, we assume $\rho^{\Gamma}$ is positive semidefinite and in this case this matrix represents a state. It is obvious that the original $\rho$ will be separable if $\rho^{\Gamma}$ is, so we will focus on the partially transposed matrix, due to its direct-sum decomposition. 

The total Hilbert space $\mathcal{H}$ has a direct-sum decomposition $\mathcal{H}=\mathcal{H}_1\oplus \mathcal{H}_2$, where $\mathcal{H}_1$ and $\mathcal{H}_2$ are the subspaces spanning $X$ and $\tilde{\rho}_{ss}$, respectively. Remembering Horodecki´s result \cite{horodecki1}, $\rho^{\Gamma}$ is separable  if and only if $I\otimes\Lambda(\rho^{\Gamma})$ is positive for any Positive but not Completely Positive (PNCP) map $\Lambda$. But as these maps are linear and since the state space is decomposed as $H_1\oplus H_2$, the PNCP maps in this case will be all of the form $\Lambda=\Lambda_1\oplus\Lambda_2$, where $\Lambda_i$ is a PNCP map acting in $H_i$. It is straightforward that $I_{A'}\otimes\Lambda_2[\tilde{\rho}_{ss}]$ is a positive operator for any PNCP $\Lambda_2$, because $\tilde{\rho}_{ss}$ is separable. 

The curious feature here is that $\mathcal{H}_1$ is in general \textit{not} a tensor product space and so it is difficult to talk about complete positivity, since the meaning of such a concept may be obscure in this situation. However, for a subspace $\mathcal{H}_1$ with reasonable low-dimension one could say that all PNCP maps are of the form $\Lambda_1=\Lambda^{CP}_a+\Lambda^{CP}_bT$, given that for a tensor product space with dimension less than $6$ all PNCP maps are of this form \cite{horodecki1,stormer}. With reasonable low-dimension we mean that there is a tensor product space with dimension less than $6$ which contains $\mathcal{H}_1$ as a subspace. In this case, $\Lambda_1$ would be seen as the restriction of PNCP maps $\Lambda^{CP}_a+\Lambda^{CP}_bT$ - all PNCP maps are of this form in this context - to the subspace $\mathcal{H}_1$. Now, as $(\rho^{\Gamma})^{\Gamma}=\rho$ is positive, we have that $I\otimes\Lambda_1(\rho^{\Gamma})$ is positive as well, for all PNCP maps $\Lambda_1$, which implies that $\rho^{\Gamma}$ is separable and hence $\rho$ is separable, i.e., $\rho$ is separable if and only if PPT. 

With this reasoning, we can construct yet several subfamilies of PPT-classifiable states. For example, take the qubit-quatrit state in the SCB given by
\begin{eqnarray*}
\rho = \left(\begin{array}{c c c c | c c c c}{x_{00}}&{0}&{0}&{0}&{0}&{0}&{0}&{0}\\{0}&{a}&{c}&{0}&{0}&{0}&{0}&{0}\\{0}&{c^*}&{b}&{0}&{0}&{0}&{0}&{0}\\ {0}&{0}&{0}&{y}&{x_{01}}&{0}&{0}&{0}\\ \hline {0}&{0}&{0}&{x_{01}^*}&{z}&{0}&{0}&{0}\\{0}&{0}&{0}&{0}&{0}&{d}&{f}&{0}\\{0}&{0}&{0}&{0}&{0}&{f^*}&{e}&{0}\\{0}&{0}&{0}&{0}&{0}&{0}&{0}&{x_{11}}\end{array}\right)
\end{eqnarray*}
and so the partial tranposed matrix is
\begin{eqnarray*}
\rho^{\Gamma} &=& \left(\begin{array}{c c c c | c c c c}{x_{00}}&{0}&{0}&{0}&{0}&{0}&{0}&{x_{01}}\\{0}&{a}&{c}&{0}&{0}&{0}&{0}&{0}\\{0}&{c^*}&{b}&{0}&{0}&{0}&{0}&{0}\\ {0}&{0}&{0}&{y}&{0}&{0}&{0}&{0}\\ \hline {0}&{0}&{0}&{0}&{z}&{0}&{0}&{0}\\{0}&{0}&{0}&{0}&{0}&{d}&{f}&{0}\\{0}&{0}&{0}&{0}&{0}&{f^*}&{e}&{0}\\{x_{01}^*}&{0}&{0}&{0}&{0}&{0}&{0}&{x_{11}}\end{array}\right) \\ &=& \left(\begin{array}{c c c c}{x_{00}}&{0}&{0}&{x_{01}}\\{0}&{y}&{0}&{0}\\{0}&{0}&{z}&{0}\\{x_{01}^*}&{0}&{0}&{x_{11}}\end{array}\right)\oplus\left(\begin{array}{c c c c}{a}&{c}&{0}&{0}\\{c^*}&{b}&{0}&{0}\\{0}&{0}&{d}&{f}\\{0}&{0}&{f^*}&{e}\end{array}\right)
\end{eqnarray*}
The first matrix support is the subspace $\mathcal{H}_{A'}\otimes\mathcal{H}_{B'}$, with $\mathcal{H}_{A'}= span\{|0\rangle,|1\rangle\}$ and $\mathcal{H}_{B'}= span\{|0\rangle,|3\rangle\}$. The second matrix represents a simply separable (unnormalized) state. %, whose support is the subspace $\mathcal{H}_{A''}\otimes\mathcal{H}_{B''}$, with $\mathcal{H}_{A''}= span\{|0\rangle,|1\rangle\}$ and $\mathcal{H}_{B'}= span\{|0\rangle,|3\rangle\}$.
 As the subspace $\mathcal{H}_{A'}\otimes\mathcal{H}_{B'}$ is four-dimensional, all PNCP maps are of the form $\Lambda^{CP}_a+\Lambda^{CP}_bT$ and by the above reasoning, the state is separable if and only if PPT. The extension to higher dimensions should be clear.

One can also apply known methods to find bound entangled PPT states restricting the search to the subspace where $X$ acts: if one proves that (\ref{ptrans}), assumed positive, is entangled, then $\rho$ will be as well and we will have a bound entangled state.     

\section{Conclusions}

We constructed novel families of bipartite states for arbitrary Hilbert space dimensions whose negative eigenvalues of the partially transposed density matrix are the negative eigenvalues of a $d_B\times d_B$ submatrix. Using this property, we presented subfamilies in which the PPT criterion is both necessary and sufficient, using the result of Proposition 1 as a major step in derivations. We also proposed a qubit-qudit novel family whose eigenvalues of the partially transposed density matrix are the same as the two-qubit and qubit-qutrit cases, irrespective of the growing dimensions of the ``core" blocks. Some nontrivial decompositions of the total Hilbert space into direct sums appeared naturally in discussions, the meaning of complete-positivity being obscure. A full mathematical and physical understanding of such situation is highly desirable. The resemblance to the Ansatz states used in \cite{verstraete} is very curious and we hope that some families proposed could be used in the same way in numerical analysis of entanglement.   If any practical implementation of some of these families in experiments is done in the future, it is immediate from our discussion that the number of resources required to detect and quantify entanglement (by Negativity) is much less than the one required for full state reconstruction. We see that even for the states considered, theoretical discussions about the partial transposition operation are not trivial and the results presented hope to shed some light in the issue. We are led then to the important question: what is in general the set of states classifiable by PPT criterion? We believe that the structure presented here combined with the reasoning of \cite{kossakowski} may bring some important results in that direction.

\section{Acknowledgments}

We are grateful to D. Chrúscínski for bringing his work on circulant states with positive partial transposition to our knowledge. We thank CAPES, and CNPQ and FAPESP, through the INCT-IQ program for financial support.

\section*{Appendix A - Direct sums and block diagonal representation of operators}
Direct sums of matrices representing operators are usually understood as block diagonal matrices. However, a different ordering of the basis of the vector space where the matrix acts gives rise to block structures differing from the usual block diagonal form. But the direct sum structure of the \textit{operator} defined by the matrix is not affected by a basis reordering. To start with a simple example, let us consider a block diagonal $4\times 4$ matrix expressed in an ordered basis $\{e_1,e_2,e_3,e_4\}$, the basis of the vector space where it acts, which we call here $\mathcal{V}$:
\begin{eqnarray}
M=\left(\begin{array}{c c c c}
{a}&{b}&{0}&{0}\\
{c}&{d}&{0}&{0}\\
{0}&{0}&{e}&{f}\\
{0}&{0}&{g}&{h}
\end{array}\right)
\end{eqnarray}
Calling
\begin{eqnarray}
M_a=\left(\begin{array}{c c c c}
{a}&{b}\\
{c}&{d}
\end{array}\right); \ \ \ \ M_b=\left(\begin{array}{c c c c}
{e}&{f}\\
{g}&{h}
\end{array}\right)
\end{eqnarray}
then we can write
\begin{eqnarray}
M=M_a\oplus M_b\label{trivial}
\end{eqnarray}
However, the direct sum symbol $\oplus$ means that the operators defined by matrices $M_a$ and $M_b$ act respectively only in subspaces $\mathcal{V}_a=span\{e_1,e_2\}$ and $\mathcal{V}_b=span\{e_3,e_4\}$ of the vector space $\mathcal{V}$. More precisely, the vector space can be decomposed as a direct sum $\mathcal{V}=\mathcal{V}_a\oplus\mathcal{V}_b$ and what (\ref{trivial}) says is that the subspaces $\mathcal{V}_a$ and $\mathcal{V}_b$ are invariant by the action of operator $M$; due to this we have the induced decomposition $M=M_a\oplus M_b$.

It is clear that the invariance of $\mathcal{V}_a$ and $\mathcal{V}_b$ by $M$ is independent of the basis ordering. If we adopt a diverse ordering, for example, $\{e_1,e_3,e_2,e_4\}$, the operator $M$ still has a diret sum structure $M=M_a\oplus M_b$. But in this new ordering, the matrix which represents $M$ no longer has a block diagonal expression, but instead:
\begin{eqnarray}
M=\left(\begin{array}{c c c c}
{a}&{0}&{b}&{0}\\
{0}&{e}&{0}&{f}\\
{c}&{0}&{d}&{0}\\
{0}&{g}&{0}&{h}
\end{array}\right)
\end{eqnarray}
In general, if a vector space $\mathcal{V}$ has a direct sum decompostion $\mathcal{V}=\bigoplus_i\mathcal{V}_i$, an operator $M$ that leaves the subspaces $\mathcal{V}_i$ invariant will have the direct sum structure $M=\bigoplus_i M_i$, where operators $M_i$ act only in subspaces $\mathcal{V}_i$ respectively. If the basis is ordered according to the subspaces $\mathcal{V}_i$, that is $\{\mathcal{B}_1, \mathcal{B}_2,..\}$, then, the matrix that represents $M$ will have a block diagonal structure. As we have many different orderings possible, clearly the matrix will not be block diagonal in general. But the decompostion $M=\bigoplus_i M_i$ is independent of this, of course.
 
Considering these remarks, we will now see how the first example of state presented in the paper, (\ref{m1}), is affected by a basis reordering.
 Adopting the following ordering of the basis,
\begin{widetext}
\begin{eqnarray*} \{|00\rangle,|11\rangle,|22\rangle,|33\rangle,|01\rangle,|02\rangle,|03\rangle,|10\rangle,|12\rangle,|13\rangle,|20\rangle,|21\rangle,|22\rangle,|30\rangle,|31\rangle,|32\rangle\}, 
\end{eqnarray*}the state (\ref{m1}) is expressed as
\begin{eqnarray}
\rho = \left(\begin{array}{c c c c | c c c c | c c c c | c c c c}
          {x_{00}}&{0}&{0}&{0}&{0}&{0}&{0}&{0}&{0}&{0}&{0}&{0}&{0}&{0}&{0}&{0} 
\\        {0}&{x_{11}}&{0}&{0}&{0}&{0}&{0}&{0}&{0}&{0}&{0}&{0}&{0}&{0}&{0}&{0} 
\\        {0}&{0}&{x_{22}}&{0}&{0}&{0}&{0}&{0}&{0}&{0}&{0}&{0}&{0}&{0}&{0}&{0}
\\        {0}&{0}&{0}&{x_{33}}&{0}&{0}&{0}&{0}&{0}&{0}&{0}&{0}&{0}&{0}&{0}&{0}
\\ \hline {0}&{0}&{0}&{0}&{a_1}&{a_4}&{a_6}&{x_{01}}&{0}&{0}&{0}&{0}&{0}&{0}&{0}&{0}
\\        {0}&{0}&{0}&{0}&{a_4^*}&{a_2}&{a_5}&{0}&{0}&{0}&{x_{02}}&{0}&{0}&{0}&{0}&{0}
\\        {0}&{0}&{0}&{0}&{a_6^*}&{a_5^*}&{a_3}&{0}&{0}&{0}&{0}&{0}&{0}&{x_{03}}&{0}&{0}
\\        {0}&{0}&{0}&{0}&{x_{01}^*}&{0}&{0}&{b_1}&{0}&{0}&{0}&{0}&{0}&{0}&{0}&{0}
\\ \hline {0}&{0}&{0}&{0}&{0}&{0}&{0}&{0}&{b_2}&{b_4}&{0}&{x_{12}}&{0}&{0}&{0}&{0}
\\        {0}&{0}&{0}&{0}&{0}&{0}&{0}&{0}&{b_4^*}&{b_3}&{0}&{0}&{0}&{0}&{x_{13}}&{0}
\\        {0}&{0}&{0}&{0}&{0}&{x_{02}^*}&{0}&{0}&{0}&{0}&{c_1}&{c_4}&{0}&{0}&{0}&{0}
\\        {0}&{0}&{0}&{0}&{0}&{0}&{0}&{0}&{x_{12}^*}&{0}&{c_4^*}&{c_2}&{0}&{0}&{0}&{0}
\\ \hline {0}&{0}&{0}&{0}&{0}&{0}&{0}&{0}&{0}&{0}&{0}&{0}&{c_3}&{0}&{0}&{x_{23}}
\\        {0}&{0}&{0}&{0}&{0}&{0}&{x_{03}^*}&{0}&{0}&{0}&{0}&{0}&{0}&{d_1}&{d_4}&{d_6}
\\        {0}&{0}&{0}&{0}&{0}&{0}&{0}&{0}&{0}&{x_{13}^*}&{0}&{0}&{0}&{d_4^*}&{d_2}&{d_5}
\\        {0}&{0}&{0}&{0}&{0}&{0}&{0}&{0}&{0}&{0}&{0}&{0}&{x_{23}^*}&{d_6^*}&{d_5^*}&{d_3}
\end{array}\right)
\end{eqnarray}
and the partially transposed matrix in this basis reads:
\begin{eqnarray}
\rho^{\Gamma} = \left(\begin{array}{c c c c | c c c c | c c c c | c c c c}
          {x_{00}}&{x_{01}}&{x_{02}}&{x_{03}}&{0}&{0}&{0}&{0}&{0}&{0}&{0}&{0}&{0}&{0}&{0}&{0} 
\\        {x_{01}^*}&{x_{11}}&{x_{12}}&{x_{13}}&{0}&{0}&{0}&{0}&{0}&{0}&{0}&{0}&{0}&{0}&{0}&{0} 
\\        {x_{02}^*}&{x_{12}^*}&{x_{22}}&{x_{23}}&{0}&{0}&{0}&{0}&{0}&{0}&{0}&{0}&{0}&{0}&{0}&{0}
\\        {x_{03}^*}&{x_{13}^*}&{x_{23}^*}&{x_{33}}&{0}&{0}&{0}&{0}&{0}&{0}&{0}&{0}&{0}&{0}&{0}&{0}
\\ \hline {0}&{0}&{0}&{0}&{a_1}&{a_4^*}&{a_6^*}&{0}&{0}&{0}&{0}&{0}&{0}&{0}&{0}&{0}
\\        {0}&{0}&{0}&{0}&{a_4}&{a_2}&{a_5^*}&{0}&{0}&{0}&{0}&{0}&{0}&{0}&{0}&{0}
\\        {0}&{0}&{0}&{0}&{a_6}&{a_5}&{a_3}&{0}&{0}&{0}&{0}&{0}&{0}&{0}&{0}&{0}
\\        {0}&{0}&{0}&{0}&{0}&{0}&{0}&{b_1}&{0}&{0}&{0}&{0}&{0}&{0}&{0}&{0}
\\ \hline {0}&{0}&{0}&{0}&{0}&{0}&{0}&{0}&{b_2}&{b_4^*}&{0}&{0}&{0}&{0}&{0}&{0}
\\        {0}&{0}&{0}&{0}&{0}&{0}&{0}&{0}&{b_4}&{b_3}&{0}&{0}&{0}&{0}&{0}&{0}
\\        {0}&{0}&{0}&{0}&{0}&{0}&{0}&{0}&{0}&{0}&{c_1}&{c_4^*}&{0}&{0}&{0}&{0}
\\        {0}&{0}&{0}&{0}&{0}&{0}&{0}&{0}&{0}&{0}&{c_4}&{c_2}&{0}&{0}&{0}&{0}
\\ \hline {0}&{0}&{0}&{0}&{0}&{0}&{0}&{0}&{0}&{0}&{0}&{0}&{c_3}&{0}&{0}&{0}
\\        {0}&{0}&{0}&{0}&{0}&{0}&{0}&{0}&{0}&{0}&{0}&{0}&{0}&{d_1}&{d_4^*}&{d_6^*}
\\        {0}&{0}&{0}&{0}&{0}&{0}&{0}&{0}&{0}&{0}&{0}&{0}&{0}&{d_4}&{d_2}&{d_5^*}
\\        {0}&{0}&{0}&{0}&{0}&{0}&{0}&{0}&{0}&{0}&{0}&{0}&{0}&{d_6}&{d_5}&{d_3}
\end{array}\right)\end{eqnarray}
having a block diagonal structure, as expected.
\end{widetext}

\end{document}